\documentclass[a4paper]{article}

\usepackage[margin=1in]{geometry}

\usepackage[T1]{fontenc}
\newcommand{\changefont}[3]{
\fontfamily{#1} \fontseries{#2} \fontshape{#3} \selectfont}

\changefont{ptm}{m}{n}

\usepackage{setspace} 
\usepackage{graphicx}    % needed for including graphics e.g. EPS, PS

\usepackage{amsfonts}
\usepackage{amssymb}
\usepackage{graphics}
\usepackage{mathrsfs}
\usepackage{color}

\newtheorem{theorem}{Theorem}[section]

\long\def\symbolfootnote[#1]#2{\begingroup%
\def\thefootnote{\fnsymbol{footnote}}\footnote[#1]{#2}\endgroup} 

\begin{document}

\begin{center}
\Large \textbf{Persistence of Chaos in Coupled Lorenz Systems}
\end{center}

\vspace{-0.3cm}
\begin{center}
\normalsize \textbf{Mehmet Onur Fen} \\
\vspace{0.2cm}
\textit{\textbf{\footnotesize Basic Sciences Unit, TED University, 06420 Ankara, Turkey}} \\
\vspace{0.1cm}
\textit{\textbf{\footnotesize E-mail: monur.fen@gmail.com, Tel: +90 312 585 0217}} \\
\vspace{0.1cm}
\end{center}
 
\vspace{0.3cm}

\begin{center}
\textbf{Abstract}
\end{center}

\vspace{-0.2cm}

\noindent\ignorespaces
The dynamics of unidirectionally coupled chaotic Lorenz systems is investigated. It is revealed that chaos is present in the response system regardless of generalized synchronization. The presence of sensitivity is theoretically proved, and the auxiliary system approach and conditional Lyapunov exponents are utilized to demonstrate the absence of synchronization. Periodic motions embedded in the chaotic attractor of the response system is demonstrated by taking advantage of a period-doubling cascade of the drive. The obtained results may shed light on the global unpredictability of the weather dynamics and can be useful for investigations concerning coupled Lorenz lasers.

\vspace{0.2cm}
 
\noindent\ignorespaces \textbf{Keywords:} Lorenz system; Persistence of chaos; Sensitivity; Period-doubling cascade; Generalized synchronization
\vspace{0.6cm}

%%%%%%%%%%%%%%%%%%%%%%%%%%%%%%%%%%%%%%%%%%%%%%%%%%%%%%%%%%%%%%%%%%%%%%%%%%%%%%%%%%%%%%%%%%%%%% 

\section{Introduction} \label{lorenz_intro}

Chaos theory, whose foundations were laid by Poincar\'{e} \cite{Poincare57}, has attracted a great deal of attention beginning with the studies of Lorenz \cite{Lorenz60,Lorenz63}. A mathematical model consisting of a system of three ordinary differential equations were introduced by Lorenz \cite{Lorenz63} in order to investigate the dynamics of the atmosphere. This model is a simplification of the one derived by Saltzman \cite{Saltzman62} which originate from the Rayleigh-B\'{e}nard convection. The demonstration of sensitivity in the Lorenz system can be considered as a milestone in the theory of dynamical systems. Nowadays, this property is considered as the main ingredient of chaos \cite{Wiggins88}.

A remarkable behavior of coupled chaotic systems is the synchronization \cite{Fujisaka83}-\cite{Rulkov95}. This concept was studied for identical systems in \cite{Pecora90} and was generalized to non-identical systems by Rulkov et al. \cite{Rulkov95}. Generalized synchronization (GS) is characterized by the existence of a transformation from the trajectories of the drive to the trajectories of the response. A necessary and sufficient condition concerning the asymptotic stability of the response system for the presence of GS was mentioned in \cite{Kocarev96}, and some numerical techniques were developed in the papers \cite{Rulkov95,Abarbanel96} for its detection.

Even though coupled chaotic systems exhibiting GS have been widely investigated in the literature, the presence of chaos in the dynamics of the response system is still questionable in the absence of GS. The main goal of the present study is the verification of the persistence of chaos in unidirectionally coupled Lorenz systems even if they are not synchronized in the generalized sense. We rigorously prove that sensitivity is a permanent feature of the response system, and we numerically demonstrate the existence of unstable periodic orbits embedded in the chaotic attractor of the response benefiting from a period-doubling cascade \cite{Franceschini80} of the drive. Conditional Lyapunov exponents \cite{Pecora90} and auxiliary system approach \cite{Abarbanel96} are utilized to show the absence of GS. Our results reveal that the chaos of the drive system does not annihilate the chaos of the response, i.e., the response remains to be unpredictable under the applied perturbation.

The usage of exogenous perturbations to generate chaos in coupled systems was proposed in the studies \cite{Akh1}-\cite{Akh6}. In particular, the paper \cite{Akhmet2015} was concerned with the extension of sensitivity and periodic motions in unidirectionally coupled Lorenz systems in which the response system is initially non-chaotic, i.e., it either admits an asymptotically stable equilibrium or an orbitally stable periodic orbit in the absence of the driving. However, in the present study, we investigate the dynamics of coupled Lorenz systems in which the response system is chaotic in the absence of the driving.

Another issue that was considered in \cite{Akhmet2015} is the global unpredictable behavior of the weather dynamics. We made an effort in \cite{Akhmet2015} to answer the question \textit{why the weather is unpredictable at each point of the Earth} on the basis of Lorenz systems. This subject was discussed by assuming that the whole atmosphere of the Earth is partitioned in a finite number of subregions such that in each of them the dynamics of the weather is governed by the Lorenz system with certain coefficients. It was further assumed that there are subregions for which the corresponding Lorenz systems admit chaos with the main ingredient as sensitivity, which means unpredictability of weather in the meteorological sense, and there are subregions in which the Lorenz systems are non-chaotic. It was demonstrated in \cite{Akhmet2015} that if a subregion with a chaotic dynamics influences another one with a non-chaotic dynamics, then the latter also becomes unpredictable. The present study takes the results obtained in \cite{Akhmet2015} a step further such that the interaction of two subregions whose dynamics are both governed by chaotic Lorenz systems lead to the persistence of unpredictability.

The rest of the paper is organized as follows. In Section \ref{LorenzSec2}, the model of coupled Lorenz systems is introduced.  Section \ref{theory} is devoted to the theoretical discussion of the sensitivity feature in the response system. Section \ref{simulations}, on the other hand, is concerned with the numerical analyses of coupled Lorenz systems for the persistence of chaos as well as the absence of GS. The existence of unstable periodic motions embedded in the chaotic attractor of the response is demonstrated in Section \ref{Lorenzzpdc}. Some concluding remarks are given in Section \ref{Lorenz_Conc}, and finally, the proof of the main theorem concerning sensitivity is provided in the Appendix.

\section{The model} \label{LorenzSec2}

Consider the following Lorenz system \cite{Lorenz63}
\begin{eqnarray}
\begin{array}{l} \label{lorenz_system}
\dot{x}_1 = -\sigma x_1 + \sigma x_2 \\
\dot{x}_2 = - x_1x_3 +rx_1 -x_2\\
\dot{x}_3 = x_1x_2-bx_3,
\end{array}
\end{eqnarray}
where $\sigma$, $r$, and $b$ are constants. 

%In system (\ref{lorenz_system}), the variable $x_1$ is proportional to the circulatory fluid flow velocity, while the variable $x_2$ is proportional to the temperature difference between the ascending and descending currents. Positive $x_1$ values indicate clockwise rotations of the fluid and negative $x_1$ values mean counterclockwise motions. The variable $x_3,$ on the other hand, is proportional to the distortion of the vertical temperature profile from linearity, a positive value indicating that the strongest gradients occur near the boundaries. The parameters $\sigma$ and $r$ are called the Prandtl and Rayleigh numbers, respectively \textbf{[Alligood et al., 1996; Lorenz, 1963; Sparrow, 1982].}

System (\ref{lorenz_system}) has a rich dynamics such that for different values of the parameters $\sigma,$ $r$ and $b,$ the system can exhibit stable periodic orbits, homoclinic explosions, period-doubling bifurcations, and chaotic attractors \cite{Sparrow82}. In the remaining parts of the paper, we suppose that the dynamics of (\ref{lorenz_system}) is chaotic, i.e., the system admits sensitivity and infinitely many unstable periodic motions embedded in the chaotic attractor. In this case, (\ref{lorenz_system}) possesses a compact invariant set $\Lambda \subset \mathbb R^3.$

Next, we take into account another Lorenz system, 
\begin{eqnarray}
\begin{array}{l} \label{nonperturbed_lorenz_system}
\dot{u}_1 = - \overline{\sigma} u_1 + \overline{\sigma} u_2 \\
\dot{u}_2 = - u_1u_3 +\overline{r} u_1 -u_2  \\
\dot{u}_3 = u_1u_2-\overline{b}u_3,
\end{array}
\end{eqnarray}
where the parameters $\overline{\sigma},$ $\overline{r}$ and $\overline{b}$ are such that system (\ref{nonperturbed_lorenz_system}) is also chaotic.  
Systems (\ref{lorenz_system}) and (\ref{nonperturbed_lorenz_system}) are, in general, non-identical, since the coefficients $\sigma,$ $r,$ $b$ and $\overline{\sigma},$ $\overline{r},$ $\overline{b}$ can be different.

We perturb (\ref{nonperturbed_lorenz_system}) with the solutions of (\ref{lorenz_system}) to set up the system
\begin{eqnarray}
\begin{array}{l} \label{perturbed_lorenz_system}
\dot{y}_1  = - \overline{\sigma} y_1 + \overline{\sigma} y_2 + g_1(x(t)) \\
\dot{y}_2  = -y_1y_3 +\overline{r} y_1 -y_2 + g_2(x(t)) \\
\dot{y}_3  = y_1y_2-\overline{b}y_3 + g_3(x(t)),
\end{array}
\end{eqnarray}
where $x(t)=(x_1(t),x_2(t),x_3(t))$ is a solution of (\ref{lorenz_system}) and $g(x)=(g_1(x),g_2(x),g_3(x))$ is a continuous function such that there exists a positive number $L_g$ satisfying $\left\|g(x) - g(\overline{x})\right\| \ge L_g \left\|x-\overline{x}\right\|$ for all $x,$ $\overline{x} \in \Lambda.$ Here, $\left\|.\right\|$  denotes the usual Euclidean norm in $\mathbb R^3.$ It is worth noting that the coupled system  $(\ref{lorenz_system})+(\ref{perturbed_lorenz_system})$ has a skew product structure. We refer to (\ref{lorenz_system}) and (\ref{perturbed_lorenz_system}) as the drive and response systems, respectively.
 
In the next section, we will demonstrate the existence of sensitivity in the dynamics of the response system.

\section{Sensitivity in the response system}  \label{theory}

Fix a point $x_0$ from the chaotic attractor of (\ref{lorenz_system}) and take a solution $x(t)$ with $x(0)=x_0.$ Since we use the solution $x(t)$ as a perturbation in (\ref{perturbed_lorenz_system}), we call it a \textit{chaotic function}. Chaotic functions may be irregular as well as regular (periodic and unstable) \cite{Lorenz63,Sparrow82}. We suppose that the response system (\ref{perturbed_lorenz_system}) possesses a compact invariant set $\mathscr{U} \subset \mathbb R^3$ for each chaotic solution $x(t)$ of (\ref{lorenz_system}). The existence of such an invariant set can be shown, for example, using Lyapunov functions \cite{Akhmet2015,Yoshizawa75}.

One of main ingredients of chaos is sensitivity \cite{Lorenz63,Wiggins88}. Let us describe this feature for both the drive and response systems.

System (\ref{lorenz_system}) is called sensitive if there exist positive numbers $\epsilon_0$ and $\Delta$ such that for an arbitrary positive number $\delta_0$ and for each chaotic solution $x(t)$ of  (\ref{lorenz_system}), there exist a chaotic solution $\overline{x}(t)$ of the same system and an interval $J \subset [0,\infty),$ with a length no less than $\Delta,$ such that $\left\|x(0)-\overline{x}(0)\right\|<\delta_0$ and $\left\|x(t)-\overline{x}(t)\right\| > \epsilon_0$ for all $t \in J.$

For a given solution $x(t)$ of (\ref{lorenz_system}), let us denote by $\phi_{x(t)}(t,y_0)$ the unique solution of (\ref{perturbed_lorenz_system}) satisfying the condition $\phi_{x(t)}(0,y_0)=y_0.$
We say that system (\ref{perturbed_lorenz_system}) is sensitive if there exist positive numbers $\epsilon_1$ and $\overline{\Delta}$ such that for an arbitrary positive number $\delta_1,$ each $y_0\in \mathscr{U}$, and a chaotic solution $x(t)$ of (\ref{lorenz_system}), there exist $y_1\in \mathscr{U},$ a chaotic solution $\overline{x}(t)$ of (\ref{lorenz_system}), and an interval $J^1 \subset [0,\infty),$ with a length no less than $\overline{\Delta},$ such that $\left\|y_0-y_1\right\|<\delta_1$ and $\left\|\phi_{x(t)}(t,y_0)-\phi_{\overline{x}(t)}(t,y_1)\right\| > \epsilon_1$ for all $t \in J^1.$

Next theorem confirms that the sensitivity feature remains persistent for (\ref{nonperturbed_lorenz_system}) when it is perturbed with the solutions of the drive system (\ref{lorenz_system}). This feature is true even if the systems (\ref{lorenz_system}) and (\ref{perturbed_lorenz_system}) are not synchronized in the generalized sense.

\begin{theorem} \label{theorem_sensitivity}
The response system (\ref{perturbed_lorenz_system}) is sensitive.
\end{theorem}

The proof of Theorem \ref{theorem_sensitivity} is provided in the Appendix. In the next section, we will demonstrate that the response system possesses chaotic motions regardless of the presence of GS.

\section{Chaotic dynamics in the absence of generalized synchronization}  \label{simulations}

Let us take into account the drive system (\ref{lorenz_system}) with the parameter values $\sigma=10,$ $r=28,$ $b=8/3$ such that the system possesses a chaotic attractor \cite{Lorenz63,Sparrow82}. Moreover, we set $\overline{\sigma}=10,$ $\overline{r}=60,$ $\overline{b}=8/3$ and $g_1(x_1, x_2, x_3) = 2.95 x_1-0.25\sin x_1,$ $g_2(x_1, x_2, x_3) = 3.06\arctan x_2,$ $g_3(x_1, x_2, x_3) = 3.12 x_3+1.75 e^{-x_3}$ in the response system (\ref{perturbed_lorenz_system}). The unperturbed Lorenz system (\ref{nonperturbed_lorenz_system}) is also chaotic with the aforementioned values of $\overline{\sigma},$ $\overline{r},$ and $\overline{b}$ \cite{Gon04,Sparrow82}. 
 
\begin{figure}[ht!] 
\centering
\includegraphics[width=9.0cm]{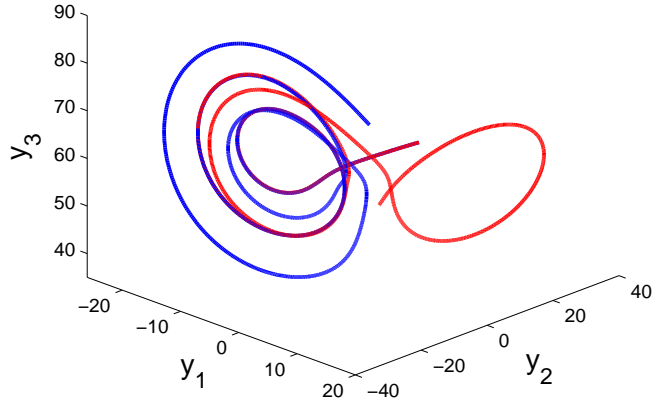}
\caption{Sensitivity in the response system (\ref{perturbed_lorenz_system}). The simulation supports the result of Theorem \ref{theorem_sensitivity} such that sensitivity is permanent in system (\ref{nonperturbed_lorenz_system}) although it is driven by the solutions of (\ref{lorenz_system}).}
\label{fig1}
\end{figure}

In order to demonstrate the presence of sensitivity in the response system (\ref{perturbed_lorenz_system}) numerically, we depict in Figure \ref{fig1} the projections of two initially nearby trajectories of the coupled system $(\ref{lorenz_system})+(\ref{perturbed_lorenz_system})$ on the $y_1-y_2-y_3$ space. In Figure \ref{fig1}, the projection of the trajectory corresponding to the initial data $x_1(0)=-8.631,$ $x_2(0)=-2.382,$ $x_3(0)=33.096,$   $y_1(0)=10.871,$ $y_2(0)=-4.558,$ $y_3(0)=70.541$ is shown in blue, and the one corresponding to the initial data $x_1(0)=-8.615,$ $x_2(0)=-2.464,$ $x_3(0)=33.067,$ $y_1(0)=10.869,$ $y_2(0)=-4.561,$ $y_3(0)=70.537$ is shown in red. The time interval $[0,2.96]$ is used in the simulation. One can observe in Figure \ref{fig1} that even if the trajectories in blue and red are initially nearby, later they diverge, and this behavior supports the result of Theorem \ref{theorem_sensitivity} such that sensitivity is present in the dynamics of (\ref{perturbed_lorenz_system}). In other words, the figure confirms that sensitivity is a permanent feature of (\ref{nonperturbed_lorenz_system}) even if it is perturbed with the solutions of (\ref{lorenz_system}).

\begin{figure}[ht!] 
\centering
\includegraphics[width=9.0cm]{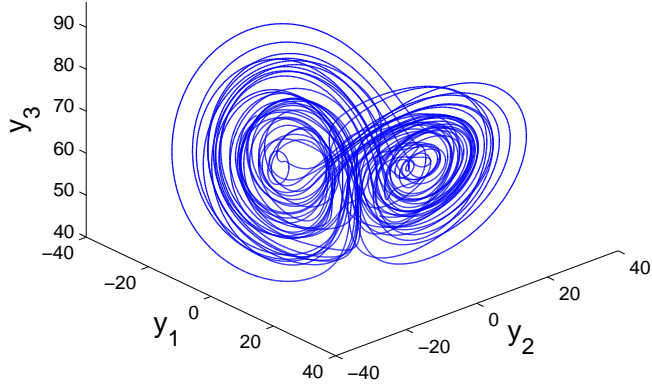}
\caption{Chaotic trajectory of system (\ref{perturbed_lorenz_system}). The figure manifests the persistence of chaos.}
\label{fig2}
\end{figure}

On the other hand, in Figure \ref{fig2}, we represent the trajectory of (\ref{perturbed_lorenz_system}) corresponding to $x_1(0)=4.43,$ $x_2(0)=-2.27,$ $x_3(0)=30.81,$ $y_1(0)=3.09,$ $y_2(0)=4.98,$ $y_3(0)=46.21.$ It is seen in Figure \ref{fig2} that the trajectory is chaotic, and this reveals the persistence of chaos in the dynamics of (\ref{perturbed_lorenz_system}).

GS \cite{Rulkov95} is said to occur in the coupled system $(\ref{lorenz_system})+(\ref{perturbed_lorenz_system})$ if there exist sets $\mathscr{X},$ $\mathscr{Y} \subset \mathbb R^3$ of initial conditions and a transformation $\psi$ defined on the chaotic attractor of (\ref{lorenz_system}) such that for all $x_0 \in \mathscr{X},$ $y_0 \in \mathscr{Y}$ the relation $\displaystyle \lim_{t\to\infty} \left\|y(t)-\psi(x(t))\right\|=0$ holds, where $x(t)$ and $y(t)$ are respectively the solutions of (\ref{lorenz_system}) and (\ref{perturbed_lorenz_system}) satisfying $x(0)=x_0$ and $y(0)=y_0.$ If GS occurs, a motion that starts on $\mathscr{X} \times \mathscr{Y}$ collapses onto a manifold $M\subset \mathscr{X}\times \mathscr{Y}$ of synchronized motions. The transformation $\psi$ is not required to exist for the transient trajectories. When $\psi$ is the identity transformation, the identical synchronization takes place \cite{Pecora90}.

It was formulated by Kocarev and Parlitz \cite{Kocarev96} that the systems (\ref{lorenz_system}) and (\ref{perturbed_lorenz_system}) are synchronized in the generalized sense if and only if for all $x_0\in \mathscr{X},$ $y_{0},$ $\overline{y}_{0}\in \mathscr{Y},$ the asymptotic stability criterion  
\begin{eqnarray*}  
\displaystyle \lim_{t\to\infty} \left\| y(t,x_0,y_{0}) - y(t,x_0,\overline{y}_{0}) \right\|=0,
\end{eqnarray*}
holds, where $y(t,x_0,y_{0}),$ $y(t,x_0,\overline{y}_{0})$ denote the solutions of (\ref{perturbed_lorenz_system}) with the initial data $y(0,x_0,y_{0})=y_{0},$ $y(0,x_0,\overline{y}_{0})=\overline{y}_{0}$ and the same solution $x(t),$ $x(0)=x_0,$ of (\ref{lorenz_system}).

A numerical technique that can be used to analyze coupled systems for the presence or absence of GS is the auxiliary system approach \cite{Abarbanel96}. We will make use of this technique for the coupled system (\ref{lorenz_system})+(\ref{perturbed_lorenz_system}). For that purpose, we consider the auxiliary system
\begin{eqnarray}
\begin{array}{l} \label{aux_lorenz_system}
\dot{z}_1 = - 10 z_1 + 10 z_2 + 2.95 x_1(t)-0.25\sin (x_1(t)) \\
\dot{z}_2 = -z_1 z_3 +60 z_1 -z_2 + 3.06\arctan (x_2(t)) \\
\dot{z}_3 = z_1 z_2- \displaystyle \frac{8}{3} z_3 + 3.12 x_3(t)+1.75 e^{-x_3(t)},
\end{array}
\end{eqnarray}
which is an identical copy of (\ref{perturbed_lorenz_system}).

\begin{figure}[ht!] 
\centering
\includegraphics[width=8.0cm]{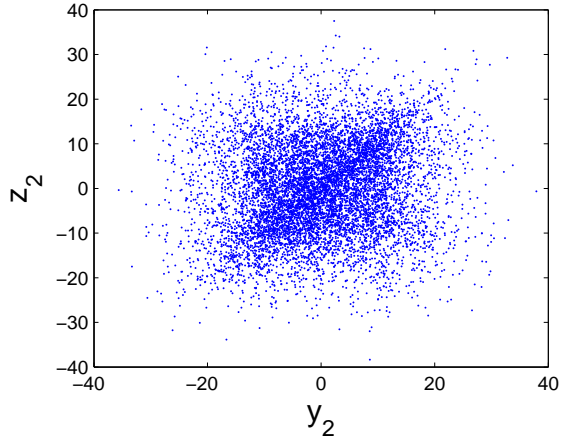}
\caption{The result of the auxiliary system approach applied to the coupled Lorenz systems (\ref{lorenz_system})+(\ref{perturbed_lorenz_system}). The figure confirms the absence of GS.}
\label{fig3}
\end{figure}

Using the initial data $x_1(0)=4.43,$ $x_2(0)=-2.27,$ $x_3(0)=30.81,$ $y_1(0)=3.09,$ $y_2(0)=4.98,$ $y_3(0)=46.21,$ $z_1(0)=7.69,$ $z_2(0)=6.25,$ $z_3(0)=52.65,$ we depict in Figure \ref{fig3} the
projection of the stroboscopic plot of the $9-$dimensional system $(\ref{lorenz_system})+(\ref{perturbed_lorenz_system})+(\ref{aux_lorenz_system})$ on the $y_2-z_2$ plane. In the simulation, the first $200$ iterations are omitted. Since the stroboscopic plot does not take place on the line $z_2=y_2,$ the systems (\ref{lorenz_system}) and (\ref{perturbed_lorenz_system}) are unsynchronized. Hence, the response system (\ref{perturbed_lorenz_system}) exhibits chaotic behavior even if GS does not occur in the systems (\ref{lorenz_system}) and (\ref{perturbed_lorenz_system}).

In order to demonstrate the absence of GS one more time by evaluating the conditional Lyapunov exponents \cite{Gon04,Pecora90,Kocarev96}, we consider the following variational system for (\ref{perturbed_lorenz_system}),
\begin{eqnarray}
\begin{array}{l} \label{cond_lyp_exp1}
\dot{\xi}_1 = -10 \xi_1 + 10 \xi_2  \\
\dot{\xi}_2 = (-y_3(t) +60) \xi_1 -\xi_2 -y_1(t) \xi_3\\
\dot{\xi}_3 = y_2(t)\xi_1 +y_1(t)\xi_2 -(8/3)\xi_3.
\end{array}
\end{eqnarray}
Making use of the solution $y(t)=(y_1(t),y_2(t),y_3(t))$ of (\ref{perturbed_lorenz_system}) corresponding to the initial conditions $x_1(0) =9.47,$ $x_2(0) =3.29,$ $x_3(0)=34.49,$ $y_1(0) =10.67,$ $y_2(0) =-8.06,$ $y_3(0)=71.89,$ the largest Lyapunov exponent of (\ref{cond_lyp_exp1}) is evaluated as $0.7693.$ In other words, the response (\ref{perturbed_lorenz_system}) possesses a positive conditional Lyapunov exponent, and this corroborates the absence of GS for the coupled systems $(\ref{lorenz_system})+(\ref{perturbed_lorenz_system}).$   

The next section is devoted to the presence of periodic motions embedded in the chaotic attractor of the response system.

\section{Periodic solutions of the response system} \label{Lorenzzpdc}

To demonstrate the existence of periodic motions embedded in the chaotic attractor of (\ref{perturbed_lorenz_system}), we will take into account (\ref{lorenz_system}) with the parameter values in such a way that the system exhibits a period-doubling cascade \cite{Feigenbaum80,Sander12}. 

Consider the drive system (\ref{lorenz_system}) in which $\sigma=10,$ $b=8/3$ and $r$ is a parameter \cite{Franceschini80,Sparrow82}. For the values of $r$ between $99.98$ and $100.795$ the system possesses two symmetric stable periodic orbits such that one of them spirals round twice in $x_1>0$ and once in $x_1<0,$ whereas another spirals round twice in $x_1<0$ and once in $x_1>0.$ The book \cite{Sparrow82} calls such periodic orbits as $x^2y$ and $y^2x,$ respectively. More precisely, $``x"$ is written every time when the orbit spirals round in $x_1>0,$ while $``y"$ is written every time when it spirals round in $x_1<0.$ As the value of the parameter $r$ decreases towards $99.98$ a period-doubling bifurcation occurs in (\ref{lorenz_system}) such that two new symmetric stable periodic orbits ($x^2yx^2y$ and $y^2xy^2x$) appear, and the previous periodic orbits lose their stability \cite{Franceschini80,Sparrow82}. According to Franceschini \cite{Franceschini80}, system (\ref{lorenz_system}) undergoes infinitely many period-doubling bifurcations at the parameter values $99.547,$ $99.529,$ $99.5255$ and so on. The sequence of bifurcation parameter values accumulates at $r_{\infty} = 99.524.$ For values of $r$ smaller than $r_{\infty}$, infinitely many unstable periodic orbits take place in the dynamics of (\ref{lorenz_system}) \cite{Franceschini80,Sparrow82}.

We say that the response (\ref{perturbed_lorenz_system}) replicates the period-doubling cascade of (\ref{lorenz_system}) if for each periodic $x(t)$, system (\ref{perturbed_lorenz_system}) possesses a periodic solution with the same period. To illustrate the replication of period-doubling cascade, let us use $\overline{\sigma}=10,$ $\overline{r}=28,$ $\overline{b}=8/3$ in (\ref{perturbed_lorenz_system}) such that the corresponding non-perturbed Lorenz system (\ref{nonperturbed_lorenz_system}) is chaotic \cite{Lorenz63,Sparrow82}. Moreover, we set $g_1(x_1,x_2,x_3)= 6.5 x_1,$ $g_2(x_1,x_2,x_3)=5.2x_2,$ $g_3(x_1,x_2,x_3)=7.1 x_3.$ One can numerically verify that the solutions of (\ref{perturbed_lorenz_system}) are ultimately bounded by a bound common for each $x(t)$. Therefore, according to Theorem $15.8$ \cite{Yoshizawa75}, the response (\ref{perturbed_lorenz_system}) replicates the period-doubling cascade of the drive (\ref{lorenz_system}). It is worth noting that the coupled system (\ref{lorenz_system})+(\ref{perturbed_lorenz_system}) possesses a period-doubling cascade as well. For the value of the parameter $r<r_{\infty},$ the instability of the infinite number of periodic solutions is ensured by Theorem \ref{theorem_sensitivity}.

\begin{figure}[ht!] 
\centering
\includegraphics[width=14.3cm]{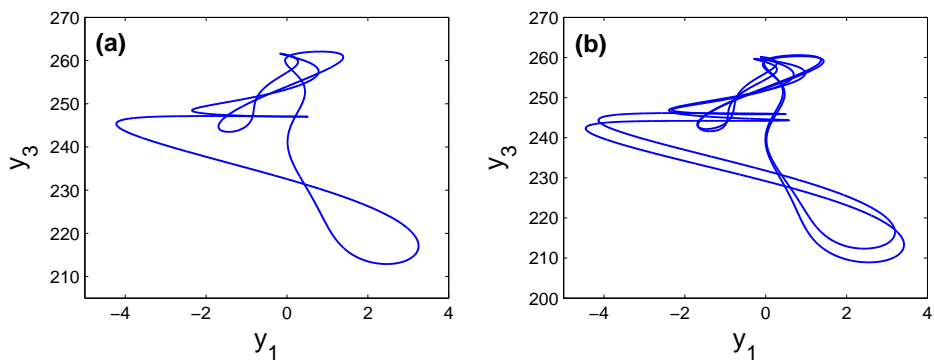}
\caption{Stable periodic orbits of the response system (\ref{perturbed_lorenz_system}). (a) Period-1 orbit corresponding to $r=100.36;$ (b) Period-2 orbit corresponding to $r=99.75.$ The pictures in (a) and (b) demonstrate the presence of periodic motions in the dynamics of (\ref{perturbed_lorenz_system}).}
\label{fig4}
\end{figure}

\begin{figure}[ht!] 
\centering
\includegraphics[width=7.2cm]{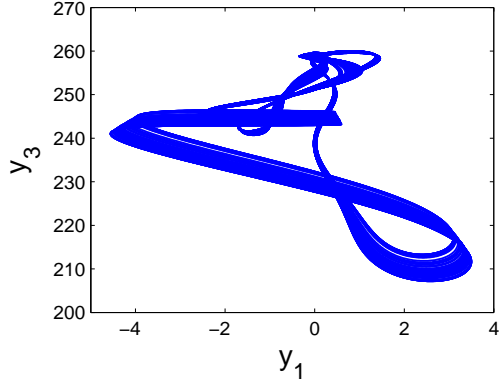}
\caption{The trajectory of the response system (\ref{perturbed_lorenz_system}) with $r=99.51.$ Chaotic behavior is observable in the figure.}
\label{fig5}
\end{figure}

Figure \ref{fig4} shows the stable periodic orbits of system (\ref{perturbed_lorenz_system}). The period-$1$ and period-$2$ orbits of (\ref{perturbed_lorenz_system}) corresponding to the $y^2x$ and $y^2xy^2x$ periodic orbits of the drive system (\ref{lorenz_system}) are depicted in Figure \ref{fig4} (a) and (b), respectively. The value $r=100.36$ is used in Figure \ref{fig4} (a), whereas $r=99.75$ is used in Figure \ref{fig4} (b). The figure reveals the presence of periodic motions in the dynamics of (\ref{perturbed_lorenz_system}). Figure \ref{fig5}, on the other hand, represents the projection of the chaotic trajectory of the coupled system $(\ref{lorenz_system})+(\ref{perturbed_lorenz_system})$ with $r=99.51$ on the $y_1-y_3$ plane. The initial data $x_1(0)=-1.15,$ $x_2(0)=3.52,$ $x_3(0)=77.01,$ $y_1(0)=0.27,$ $y_2(0)=2.17,$ $y_3(0)=254.09$ are used in the simulation. Figures \ref{fig5} manifests that (\ref{perturbed_lorenz_system}) replicates the period-doubling cascade of (\ref{lorenz_system}).

%For a fixed value of the parameter $r<99.524,$ one can numerically verify that the solutions of (\ref{perturbed_lorenz_system}) are ultimately bounded by a bound common for each solution $x(t)$ of (\ref{lorenz_system}). Therefore, the replication of the period-doubling cascade of the drive system (\ref{lorenz_system}) by the response system (\ref{perturbed_lorenz_system}) can be approved theoretically by means of Theorem $15.8$ \cite{Yoshizawa75}. 

\section{Conclusions} \label{Lorenz_Conc}

In the present study, we demonstrate the persistence of chaos in unidirectionally coupled Lorenz systems by checking for the existence of sensitivity and infinitely many unstable periodic motions. This is the first time in the literature that the presence of sensitivity in the dynamics of the response is theoretically proved regardless of GS. The obtained results certify that the applied perturbation does not suppress the chaos of the response.  We make use of conditional Lyapunov exponents \cite{Pecora90} and the auxiliary system approach \cite{Abarbanel96} to show the absence of GS in the investigated systems. It is worth noting that the results are valid for both identical and non-identical systems, i.e., the coefficients of the coupled Lorenz systems can be same or different.

One of the concepts related to our results is the global unpredictable behavior of the weather dynamics. This subject was considered in \cite{Akhmet2015} on the basis of Lorenz systems assuming that the whole atmosphere of the Earth is partitioned in a finite number of subregions such that in each of them the dynamics of the weather is governed by the Lorenz system with certain coefficients. The present paper plays a complementary role to the discussions of \cite{Akhmet2015} in such a way that the unpredictable behavior of the weather dynamics is permanent under the interaction of two subregions whose dynamics are both governed by chaotic Lorenz systems. Another concept in which Lorenz systems are encountered is the laser dynamics. It was shown by Haken \cite{Haken75} that the Lorenz model is identical with that of the single mode laser. Therefore, our results may also be used as an engineering tool to design unsynchronized chaotic Lorenz lasers \cite{Lawandy87}.

\section*{Appendix: Proof of Theorem \ref{theorem_sensitivity}}

The proof of Theorem \ref{theorem_sensitivity} is as follows.

\noindent \textbf{Proof.}
Fix an arbitrary positive number $\delta_1,$ a point $y_0\in \mathscr{U},$ and a chaotic solution $x(t)$ of (\ref{lorenz_system}). One can find $\epsilon_0>0$ and $\Delta>0$ such that for arbitrary $\delta_0>0$ both of the inequalities $\left\|x(0)-\overline{x}(0)\right\|<\delta_0$ and $\left\|x(t)-\overline{x}(t)\right\| > \epsilon_0,$ $t \in J,$ hold for some chaotic solution $\overline{x}(t)$ of (\ref{lorenz_system}) and for some interval $J \subset [0,\infty),$ whose length is not less than $\Delta.$ 

Take an arbitrary $y_1\in \mathscr{U}$ such that $\left\|y_0-y_1\right\|<\delta_1.$ For the sake of brevity, let us denote $y(t)=\phi_{x(t)}(t,y_0)$ and $\overline{y}(t)=\phi_{\overline{x}(t)}(t,y_1).$  It is worth noting that there exist positive numbers $K_0$ and $H_0$ such that $\left\|y(t)\right\|, \left\|\overline y(t)\right\| \le K_0$ for all $t\ge 0$, and $\displaystyle \sup_{t \ge 0} \left\|x(t)\right\| \leq H_0$ for each chaotic solution $x(t)$ of (\ref{lorenz_system}).
 
Our aim is to determine positive numbers $\epsilon_1,$ $\overline{\Delta}$ and an interval $J^1\subset [0,\infty)$ with length $\overline{\Delta}$ such that the inequality $\left\|y(t)-\overline{y}(t)\right\| > \epsilon_1$ holds for all $t \in J^1.$

It is clear that the collection of chaotic solutions of system (\ref{lorenz_system}) is an equicontinuous family on $[0,\infty).$  Making use of the uniform continuity of the function $\overline{g}: \mathbb R^3 \times \mathbb R^3  \to \mathbb R^3,$ defined as $\overline{g}(\nu_1,\nu_2)=g(\nu_1)-g(\nu_2),$ on the compact region 
$
\mathscr{R}=\left\{(\nu_1,\nu_2) \in \mathbb R^3 \times \mathbb R^3 : \left\|\nu_1\right\| \leq H_0, \left\|\nu_2\right\| \leq H_0 \right\} 
$
together with the equicontinuity of the collection of chaotic solutions of (\ref{lorenz_system}), one can verify that the collection $\mathscr{F}$ consisting of the functions of the form
$g_j(x_1(t))-g_j(x_2(t)),$ $j=1,2,3,$ where $x_1(t)$ and $x_2(t)$ are chaotic solutions of system (\ref{lorenz_system}), is an equicontinuous family on $[0,\infty).$

According to the equicontinuity of the family $\mathscr{F},$ one can find a positive number $\tau<\Delta,$ which is independent of $x(t)$ and $\overline{x}(t),$  such that for any $t_1,t_2\in [0,\infty)$ with $\left|t_1-t_2\right|<\tau,$ the inequality 
\begin{eqnarray} \label{sensitivity_proof_1}
\begin{array}{l}
\left| \left(g_j\left(x(t_1)\right) - g_j\left(\overline{x}(t_1)\right)  \right) - \left(g_j\left(x(t_2)\right) - g_j\left(\overline{x}(t_2)\right)  \right)   \right| <\displaystyle \frac{L_g\epsilon_0}{6}
\end{array}
\end{eqnarray}
holds for all $j=1,2,3.$

On the other hand, for each $t\in J,$ there exists an integer $j_0,$ $1 \leq j_0 \leq 3,$ which possibly depends on $t,$ such that 
\begin{eqnarray}
\begin{array}{l}
\left|g_{j_0}(x(t))-g_{j_0}(\overline{x}(t))\right|   \geq \displaystyle \frac{L_g}{3} \left\|x(t)-\overline{x}(t)\right\| \nonumber.
\end{array}
\end{eqnarray}
Otherwise, if there exists $s\in J$ such that the inequality  
\begin{eqnarray}
\begin{array}{l}
\left|g_{j} \left(x\left(s\right)  \right)-g_{j}(\overline{x}(s)) \right|< \displaystyle \frac{L_g}{3} \left\|x(s)-\overline{x}(s)\right\| \nonumber
\end{array}
\end{eqnarray}
holds for each $j=1,2,3,$ then one encounters with a contradiction since
\begin{eqnarray*}
\left\|g(x(s))-g(\overline{x}(s))  \right\|   \leq \sum_{j=1}^{3}\left| g_{j}(x(s))-g_{j}(\overline{x}(s)) \right|  < L_g \left\|x(s)-\overline{x}(s)\right\|. 
\end{eqnarray*}

Denote by $s_0$ the midpoint of the interval $J,$ and let $\displaystyle \theta=s_0-\tau/2.$ There exists an integer $j_0,$  $1 \leq j_0 \leq 3,$ such that 
\begin{eqnarray}
\begin{array}{l} \label{sensitivity_proof_2}
\left|g_{j_0}(x(s_0))-g_{j_0}(\overline{x}(s_0))\right|   \geq \displaystyle\frac{L_g}{3} \left\|x(s_0)-\overline{x}(s_0)\right\| > \displaystyle\frac{L_g\epsilon_0}{3}. 
\end{array}
\end{eqnarray}
Using the inequality (\ref{sensitivity_proof_1}) it can be verified for all $t \in \left[\theta, \theta+\tau\right]$ that
\begin{eqnarray*}
&& \left|g_{j_0}\left(x(s_0)\right) - g_{j_0}\left(\overline{x}(s_0)\right) \right| - \left|g_{j_0}\left(x(t)\right) - g_{j_0}\left(\overline{x}(t)\right) \right| \\
&& \leq \left| \left(g_{j_0}\left(x(t)\right) - g_{j_0}\left(\overline{x}(t)\right)  \right) - \left(g_{j_0}\left(x(s_0)\right) - g_{j_0}\left(\overline{x}(s_0)\right)  \right)   \right| \\
&&<\frac{L_g\epsilon_0}{6}.
\end{eqnarray*}
Therefore, by means of  (\ref{sensitivity_proof_2}), we have 
\begin{eqnarray} \label{sensitivity_proof_3}
\begin{array}{l} 
 \left|g_{j_0}\left(x(t)\right) - g_{j_0}\left(\overline{x}(t)\right) \right| > \left|g_{j_0}\left(x(s_0)\right) - g_{j_0}\left(\overline{x}(s_0)\right) \right|  - \displaystyle \frac{L_g\epsilon_0}{6} > \displaystyle \frac{L_g\epsilon_0}{6}
 \end{array}
\end{eqnarray}
for  $t\in \left[\theta, \theta+\tau\right].$

One can find numbers $s_1, s_2, s_3 \in [\theta,\theta+\tau]$ such that
\begin{eqnarray*}
\displaystyle\int^{\theta + \tau}_{\theta} \left[g(x(s))-g(\overline{x}(s))\right] ds   = \Big(  
\tau \left[g_1(x(s_1))-g_1(\overline{x}(s_1))\right],  
\tau \left[g_2(x(s_2))-g_2(\overline{x}(s_2))\right],  \\ 
\tau \left[g_3(x(s_3))-g_3(\overline{x}(s_3))\right] 
  \Big).
\end{eqnarray*}
Inequality (\ref{sensitivity_proof_3}) yields
\begin{eqnarray*}  
\Big\|\displaystyle\int^{\theta + \tau}_{\theta} \left[g(x(s))-g(\overline{x}(s))\right] ds \Big\| \geq \tau  \left|g_{j_0}(x(s_{j_0}))-g_{j_0}(\overline{x}(s_{j_0}))\right| > \displaystyle \frac{\tau  L_g \epsilon_0}{6}.
\end{eqnarray*}

Let us define the function $f:\mathbb R^3 \to \mathbb R^3$ as 
$
f(v) = ( - \overline{\sigma} v_1 + \overline{\sigma} v_2, -v_1v_3 +\overline{r} v_1 -v_2, v_1v_2-\overline{b}v_3),
$ where $v=(v_1,v_2,v_3).$ One can confirm the presence of a positive number $L_f$ such that $\left\|f(v)-f(\overline{v})\right\| \le L_f \left\|v-\overline{v}\right\|$ for all $v,$ $\overline{v} \in \mathscr{U}.$
The relation
\begin{eqnarray*}
y(t)-\overline{y}(t)  = (y(\theta)-\overline{y}(\theta)) + \displaystyle\int^{t}_{\theta}  \left[ f(y(s))-f(\overline{y}(s)) \right]   ds  + \displaystyle\int^{t}_{\theta} [g(x(s))-g(\overline{x}(s))]  ds,   ~t\in [\theta,\theta+\tau]
\end{eqnarray*}
implies that
\begin{eqnarray*} 
& \left\| y(\theta+\tau)-\overline{y}(\theta+\tau) \right\| & \geq  \Big\|\displaystyle\int^{\theta+\tau}_{\theta}  [g(x(s))-g(\overline{x}(s))] ds \Big\|  - \left\| y(\theta)-\overline{y}(\theta) \right\|\\
&& - \displaystyle\int^{\theta+\tau}_{\theta}  L_f\left\| y(s)-\overline{y}(s) \right\| ds \\
&& > \displaystyle \frac{\tau  L_g \epsilon_0}{6} - \left\| y(\theta)-\overline{y}(\theta) \right\| - \displaystyle\int^{\theta+\tau}_{\theta}  L_f\left\| y(s)-\overline{y}(s) \right\| ds.
\end{eqnarray*}
According to the last inequality we have that
\begin{eqnarray*}
& \displaystyle \max_{t\in [\theta,\theta+\tau]}\left\| y(t)-\overline{y}(t)\right\| & \geq \left\| y(\theta+\tau)-\overline{y}(\theta+\tau) \right\| \\
&& > \frac{\tau L_g \epsilon_0}{6}  - (1+ \tau L_f) \displaystyle \max_{t\in [\theta,\theta+\tau]}\left\| y(t)-\overline{y}(t) \right\|.
\end{eqnarray*}
Therefore,  
$
\displaystyle \max_{t\in [\theta,\theta+\tau]}\left\| y(t)-\overline{y}(t) \right\| > \frac{\tau L_g \epsilon_0}{6(2+\tau L_f)}.
$

Suppose that  
$
\displaystyle \max_{t \in [\theta,\theta+\tau]} \left\|y(t)-\overline{y}(t)\right\|  = \left\|y(\xi)-\overline{y}(\xi)\right\| 
$
for some $\xi \in [\theta, \theta+\tau].$ 
Define the number
\begin{displaymath}
\overline{\Delta}=\min \displaystyle \left\{ \frac{\tau}{2}, \frac{\tau L_g \epsilon_0}{24(K_0L_f+M_g)(2+\tau L_f)}   \right\},
\end{displaymath}
where $M_g = \displaystyle \sup_{x \in \Lambda} \left\|g(x)\right\|,$ and let
\begin{displaymath}
\theta^1=\left\{\begin{array}{ll} \xi, & ~\textrm{if}~  \xi \leq \theta + \tau/2   \\
\xi - \overline{\Delta}, & ~\textrm{if}~  \xi > \theta + \tau/2  \\
\end{array} \right. .\nonumber
\end{displaymath}

For $t\in [\theta^1, \theta^1+\overline{\Delta}],$ by favour of the equation
\begin{eqnarray*}
y(t)-\overline{y}(t) = (y(\xi)-\overline{y}(\xi)) + \displaystyle\int^{t}_{\xi} \left[f(y(s))-f(\overline{y}(s))\right] ds  + \displaystyle\int^{t}_{\xi} [g(x(s))-g(\overline{x}(s))]  ds,
\end{eqnarray*}
one can obtain that
\begin{eqnarray*} 
& \left\|y(t)-\overline{y}(t)\right\| & \geq  \left\|y(\xi)-\overline{y}(\xi)\right\|  - \left|  \displaystyle\int^{t}_{\xi} L_f \left\| y(s)-\overline{y}(s)\right\| ds  \right|\\ 
&& -  \left|  \displaystyle\int^{t}_{\xi} \left\|  g(x(s))-g(\overline{x}(s))  \right\| ds  \right| \\
&& > \displaystyle\frac{ \tau L_g \epsilon_0}{6(2+\tau L_f)} -2\overline{\Delta} \left(K_0L_f+M_g \right) \\
&& \geq \displaystyle\frac{\tau L_g \epsilon_0}{12(2+\tau L_f)}.
\end{eqnarray*}

The length of the interval $J^1=[\theta^1, \theta^1+\overline{\Delta}]$ does not depend on $x(t),$ $\overline{x}(t),$ and for $t \in J^1$ the inequality
$
\left\|y(t)-\overline{y}(t)\right\| > \epsilon_1
$ 
holds, where $\epsilon_1=\displaystyle \frac{\tau L_g \epsilon_0}{12(2+\tau L_f)}.$ Consequently, the response system (\ref{perturbed_lorenz_system}) is sensitive. $\square$

%\section*{References} 

\end{document}